\documentclass[usenatbib]{mn2e}
\usepackage{graphicx}
\usepackage{bm}
\usepackage{amssymb} % (adds particular binary math symbols) 
\usepackage{amsmath}%,tensor}
\def\sigmaB{\sigma_{\rm B}}
\def\sigmaR{\sigma_{\rm R}}
\def\vecX{\bm{X}}
\def\vecY{\bm{Y}}
\def\matB{\mathsf B}
\def\matH{\mathsf H}
\def\matR{\mathsf R}
\def\matBB{\bm{\mathsf B}}
\def\matHH{\bm{\mathsf H}}
\def\matRR{\bm{\mathsf R}}
\def\matWW{\bm{\mathsf W}}
\def\wf{w}

\title{Data assimilation for stratified convection}
\author[A. Svedin, et al.]{
Andreas Svedin$^1$\thanks{E-mail: aos2112@columbia.edu},
Milena C.\ Cu\'ellar$^2$
and
Axel Brandenburg$^{3,4}$
\vspace{2mm}
\\
$^1$Astronomy Department, Columbia University, New York, NY 10027, USA\\
$^2$CUNY - Bronx Community College, 2155 University Avenue  Bronx,
NY 10453, USA\\
$^3$Nordita, KTH Royal Institute of Technology and Stockholm University, Roslagstullsbacken 23,
SE-10691 Stockholm, Sweden\\
$^4$Department of Astronomy, Stockholm University, SE-10691 Stockholm, Sweden
}
\begin{document}
\maketitle
\begin{abstract}
We show how the 3DVAR data assimilation methodology can
be used in the astrophysical context 
of a two-dimensional convection
flow. 
We study the way this variational 
approach finds best estimates of 
the current state of the flow from a 
weighted average of model states and observations. 
We use numerical simulations to generate synthetic
observations of 
a vertical two-dimensional slice of the 
outer part of the solar convection zone 
for varying noise levels and
implement 3DVAR when the 
covariance matrices are scalar.  
Our simulation results demonstrate 
the capability of 3DVAR to produce
error estimates of system states 
between up to tree orders of
magnitude below the original 
noise level present in the observations.  
This work exemplifies the importance 
of applying data assimilation
techniques in simulations 
of the stratified convection.  

\end{abstract}

\section{Introduction}
%%%%%%%%%%%%%%%
When using models to describe the 
temporal evolution of observed
complex systems we are confronted 
with a number of challenges.
An immediate difficulty in dealing with this
question is that we generally do not know in 
all detail the current state of the
system or the initial condition that
is to be used.  
Lacking such information prevents us from 
keeping a model-based simulation in
step with the behavior of the observed system.  

\emph{Data assimilation} techniques offer means to address
such challenges for complex systems by keeping a
computer simulation (i.e.\ model) in 
synchronization with observations of
the system it represents. 
It provides a general framework for 
simultaneously comparing, combining, and 
evaluating observations of physical systems and
output from computer simulations.
The methods used in data assimilation have been 
developed over several decades, primarily in meteorology 
and oceanography for the prediction of the future behavior.

Data assimilation is used daily in operational
weather prediction \citep{BGK81}, 
in climate forecast \citep{PH06} 
and it was even used to correct the path of the Apollo spacecraft
during the first moon landings 
\citep{cipra1993engineers}.
There is a large and growing body of literature
including several monographs 
\citep{daley1993,kalnay2003,wunsch2006}
and work discussing its theoretical foundations
\citep{L81,lorenc1986,LdT86,ghil1989}.
Astrophysical data assimilation has recently been discussed by
\citep{Brun2007}, both
in the context of space weather and in solar cycle prediction
\citep{Dikpati2007,choudhuri2007,kitiashvili2008}
as well as in the context of dynamo models \citep{Jouve2011}.

Here we focus on the three-dimensional 
variational (3DVAR) data
assimilation technique, 
also known as sequential approach
\citep{daley1993}, which produces 
updates of the current state of a
model simulation at times when 
system observations are available.  
Propagation of model states between 
times when the system is observed
are free simulations of the model 
initiated at the latest state
estimate.  
An extension of 3DVAR to implicitly 
incorporate dynamical information is known as 
four-dimensional variational (4DVAR)
data assimilation. 
3DVAR dynamically evolves the mean state
wheres 4DVAR also evolution other statistical
properties of the model dynamics.

State estimates produced by 3DVAR are 
optimal provided that the model
is linear and the uncertainties are Gaussian. 
In other words, 3DVAR states are
Best Linear Unbiased Estimates, where \emph{best} 
and \emph{optimal} refer to the 
lowest possible mean squared  error of
the estimate \citep{K60,T97}. 

For nonlinear models, error statistics 
may become non-Gaussian
even when the initial distribution 
is Normal and 3DVAR (or 4DVAR)
estimates are not longer unbiased. 
In this case, data assimilation techniques 
are challenged by the fact
that actual applications are 
typically based on nonlinear processes
\citep{PVT96}. 
Specifically, that the states exhibited by 
real systems under observation will
diverge from those predicted by 
a model simulation is clear and
this is principally owing to 
two causes (Palmer \& Hagedorn 2006):
observational error and sensitivity to initial conditions.  
The first of these is a result of
what may be called noise.  
Since its statistical character
may not be known, we may need to make some assumptions
about its properties.  
The second source of error occurs on many complex
systems and is referred to as chaotic behavior.  
This has been
known for some time, but only in recent decades serious
progress in its understanding has been possible.
Sensitivity of the model to initial conditions limits
how far into the future predictions can be made \citep{L93}.

Among other challenges present for data assimilation of
nonlinear model states like 
the mismatch between spatial locations of  
observations and grid positions of the model, and 
the unpaired model variables to observed quantities make
the study towards more effective 
data assimilation techniques an
important and demanding area of research. 
Despite these challenges and open questions, 
3DVAR is widely used in the 
oceanographic and meteorological communities,
and would make a good candidate 
method to be explored in the context of astrophysical flows.

\section{Stratified Convection Model}\label{sec:Strat}

We are motivated to use data assimilation 
techniques in the context of
stratified convection as a path 
to obtain predictions of solar
subsurface weather events, i.e.\
the flow structure beneath the surface.
Anticipating the possibility of violent
events on the solar surface 
such as coronal mass ejections 
that affect the space weather 
and the dynamics of the Earth's 
magnetosphere is important,
\citep[see][]{Ilonidis2011S}.
The idea is to use a model of solar subsurface convection,
ultimately involving the magnetic 
fields that give rise to surface activity
such as coronal mass ejections, 
although current attempts in that
direction are still at a preliminary stage \citep{WBM11}.
However, once such models are able to 
reproduce sufficient details
of solar activity, 
it will be important to synchronize the model
with daily observations to be 
able to use it for predictions.

As a proof of concept, we design 
a data assimilation experiment to test the 
implementation of 3DVAR for the {\sc Pencil Code},
a public domain code of high-order 
(sixth order in space and third order in time) for
solving the hydrodynamic 
equations \citep{BD02}\footnote{\texttt{
http://pencil-code.googlecode.com/}}.
We consider here a simple two-dimensional 
convection model 
representing the turbulent flows of 
stars with outer convection zones.
In our experiment, \emph{synthetic observations} are 
generated by adding noise to the output from our model.
These observations are then processed 
by 3DVAR to produce an \emph{analysis}.
An analysis is an estimation of the 
\emph{unknown} state of a system in 
terms of model variables
\citep{lorenc1986,T97}.

Our implementation is general and 
can in the future be used 
for other problems that can 
be addressed with the {\sc Pencil Code}.
In this work we assume the model to be 
\textit{ideal} and reproduces the same features present
in the observations.
In real world applications, the models are far from ideal, and
imperfections and uncertainties related to the model are always
present. 
Ideally, we would like to be able to 
account for some portion of those
unknowns by using data assimilation techniques.

We use the sample \texttt{2d/conv-slab-MLT}
of the {\sc Pencil Code} (revision r14696 and later).
This sample simulates a vertical 
two-dimensional slice of the outer part of 
a stellar convection zone. 
In particular, we use it to simulate convection
at low resolution, $64\times64$, 
at a Rayleigh number of
$8\times10^5$ \citep{DBNS05}, 
and a Reynolds number of approximately
$30$. 
The basic setup is similar to that 
described in \citep{BCNS05}
and other models before them \citep{HTM86,Betal96}, 
consisting of a convectively unstable 
layer sandwiched between two stable layers. 

The simulated vertical two-dimensional slice 
of the outer part of the solar
convection zone has a  
mean field velocity of  $u_{\rm rms}=0.08$, 
the wavenumber of the energy-carrying eddies 
is $k_{\rm f}=2\pi d$ for a depth of
the unstable layer $d=1$, 
therefore the correlation time
$\tau_{\rm cor}=(u_{\rm rms}k_{\rm f})^{-1}$ is approximately 2. 

Starting from an initial velocity field of perturbations 
with an amplitude of $3\times10^{-4}$ times 
the average sound speed, convective
motion is generated without 
having to introduce any stochastic
elements.  
This model is chosen to illustrate 3DVAR
in an astrophysical context for its sufficiently 
complex behavior without having any stochastic elements.

%%%%%%%%%%%%%%%
\section{Data assimilation setup}
%%%%%%%%%%%%%%%
\label{sec:expsetup}

The 3DVAR scheme was developed in the meteorological 
community to improve model-based weather prediction 
in spite of observational and modeling uncertainties.
It was formulated in a unified Bayesian 
framework by \citep{lorenc1986}.
3DVAR produces updates of the current state 
of the system at times
when observations are available, which in turn  
can be used as a new initial condition to be 
propagated forward to the time
when the next observation is available. 
We can use 3DVAR as a black
box along with a low resolution simulation to 
assimilate many data points at low computational cost on a
laptop computer. 
For example, a typical model run for a $64\times64$
two-dimensional convection field over 
a time interval of 300 time
units takes about
15 minutes on a laptop computer.

3DVAR minimizes the sum of the 
squared differences between 
both the model background state and the observations
to find a solution that is a 
compromise between these two estimates of the true state.

It is important to realize that 
in real problems, the true state 
is available only through noisy
observations of the system. 
Therefore, it is impossible to tell how 
close our model output is to
current and future true states of the system. 
In our \textit{twin-experiment}
 \citep{bengtsson1981dynamic}
we select two different initial
 fields to run the {\sc Pencil Code}
simulation.  
One of these initial fields
 represents the unknown true initial
state of the system.
The other initial field represents
what might be, in practice, a good approximation 
or guess, of the initial state of the system.

As mentioned before, we are set at
 the ideal case where there is no model 
uncertainty and the only source
 of uncertainty is in the observations. 
In this way, we can assess how
 far/close the model state is to
the true state of the system. 
The key is to generate a known
 true state against which the 
estimated state obtained via
 data assimilation can be verified.

The experiment is setup as follows: 
the initial field chosen to
represent the true initial state of 
the system, initializes a model run
considered to be the original  
state of the system to be used as 
reference trajectory or \emph{control}.
The other initial field is used to 
initialize two different runs: one
a \emph{free} model run and other 
that will become the assimilated
trajectory or \emph{analysis}. 
The analysis is a collection of
segments of model trajectories initialized 
at the 3DVAR corrections
made at all times where the observations are available.
The model (analysis) states at the 
time when an observation is
available are known as the \emph{background states}.

Comparison of the free and the control run 
gives us a measure of the
sensitivity to initial conditions of our model, 
i.e.\ it shows how similar 
initial conditions diverge in time.

Similarly, comparing the free run and the analysis
represents the effect of the data assimilation 
procedure over a trajectory starting at the same 
initial field. 
If the assimilation of the second set 
of initial conditions is effective
it will bring the analysis ``closer'' to the control run, 
and further away from the free run.

We generate synthetic observations 
by adding independent and identically distributed noise
to the horizontal velocity field in all
grid-points to the control run. 
These synthetic data are considered to 
be our experimental observations
which in turn will be 
used to update the analysis. 
As is explained in the next section, 3DVAR requires both a
model and observation state to find 
update the analysis at each given
time in the assimilation window.

%%%%%%%%%%%%%%%%%%
\section{3DVAR and the weight factor}\label{sec:practice}
%%%%%%%%%%%%%%%%%%
The 3DVAR technique finds a model state $\vecX$
that agrees with the current state of 
the system and the information
available in the observations and the model.
Specifically, we minimize the 
weighted average of the residues for both observations 
$\vecY_0$ and model
states $\vecX_b$ at time $t$ to find an optimal solution.
This is expressed in the 3DVAR cost function \citep{lorenc1986} 
\begin{equation}
\begin{aligned}
J(\vecX)=&\frac{1}{2}\left[ \vecX-\vecX_b \right]^T \matBB^{-1}\left[
\vecX-\vecX_b\right]\\
+&\frac{1}{2}\left[\vecY_0 - \matHH(\vecX)\right]^T 
\matRR^{-1}\left[\vecY_0 - \matHH(\vecX)\right], \label{cost3d}
\end{aligned}
\end{equation}
where $\vecX_b$ is the model state--traditionally called the 
\emph{background state}, $\vecY_0$ 
is the observed state, 
$\matBB$ is the background covariance matrix,
$\matRR$ is the observational covariance matrix,
and $\matHH$ is the observation operator.

The analysis is $\vecX_a$, minimizes 
$J(\vecX)$ and corresponds to
the best estimate of the current state of the system. 
At that time, the model is integrated 
using as initial condition the
analysis up to the next time 
an observation is available. 
 
Synthetic observations are denoted by $\vecY_0$, 
these observations
contain noise of amplitude, $\sigmaR$, 
proportional to the maximum
amplitude of the full 2D vertical velocity field.
For example, a noise level of 1\% corresponds 
to $\sigmaR=0.01$
for a normalized field or $\sigmaR=6\times10^{-3}$ 
for an unnormalized field.

The selection and construction of the observational 
and background 
covariance matrices ($\matRR$ y $\matBB$) 
is of great interest in data
assimilation \citep{bannister2008a,bannister2008b}.
In our case, the observational noise is 
not correlated in space,
and we neglect spatial correlations between
model states, 
making the off-diagonal components
of the $\matBB$ and $\matRR$ matrices vanish.
By doing this, we can set these matrices to be scalars,
$\matR_{ij}=\delta_{ij}\sigmaR^2$ and   
$\matB_{ij}=\delta_{ij}\sigmaB^2$. 
Without these spatial correlations 3DVAR generates an
analysis that is, in general, 
less smooth over the two-dimensional domain.
In more sophisticated formulations of equation (\ref{cost3d}), 
the form of $\matBB$ can also include physical constraints to 
processes not resolved in the model. 
Literature in this area is extensive, especially
in oceanography and meteorology \citep{
dobricic2008}. 

In turn, we set the observation operator to be
$\matH_{ij}=\delta_{ij}$.  
This means that we assume observations cover all
grid-points in the model domain, i.e.\
the observables and model variables 
belong to the same space. 
In other words, system and model are the same.
In more realistic implementations of 3DVAR, 
$\matHH$ is typically a computer algorithm 
that cannot be expressed explicitly as a matrix
due to its nonlinear nature \citep{dobricic2008}. 
For example the transformation between observables
and variable might require the modeling of 
dynamical processes or making averages.  

After these assumptions 
over the matrices in (\ref{cost3d}), we can
translate the cost function to:
\begin{equation}
J(\vecX)= \wf \left(\vecX-\vecX_b \right)^2 +
\left(\vecX-\vecY_0\right)^2, \label{cost3d2}
\end{equation}
where $\wf$ is the ratio of the scalar variances
corresponding to the observed and background states
\begin{equation}
\wf=(\sigmaR/\sigmaB)^2.
\end{equation}
In this setting, to find the state 
vector $\vecX_a$ that minimizes
(\ref{cost3d2}), 
we use {\sc Powell} minimization \citep{press2007numerical}.

The coefficient $\wf$ in equation 
(\ref{cost3d2}) behaves as a
weight in the optimization process 
and will be referred to as the
\emph{weight factor}. 

Solving $\bm{\nabla} J(\vecX)=\bm{0}$ yields 
\begin{equation}
w(\vecX-\vecX_b)+(\vecX-\vecY_0)=\bm{0}, 
\end{equation}
and 
the optimal state of the model that 
represents the systems given the
current observations, or %
analysis $\vecX_a\equiv\vecX$,
is given by: %]
\begin{equation}
\vecX_a=\frac{1}{\wf+1}\vecY_0+\frac{\wf}{\wf+1}\vecX_b.
\label{eq:xa}
\end{equation}
Equation (\ref{eq:xa}) unveils how 
the contribution of the background states $\vecX_b$
and observations $\vecY_0$
affects 
the analysis 
$\vecX_a$ in terms of the weight factor $\wf$. 

\begin{figure}
\includegraphics[width=\columnwidth]{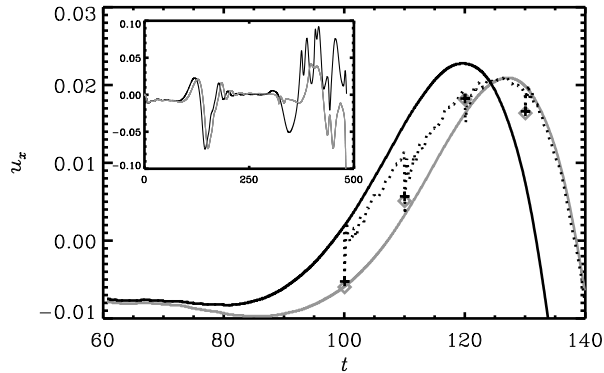}
\caption{Temporal evolution of the horizontal 
velocity at a certain midpoint
of a 2D convection field for $t\in[0,500]$. 
Control, free run, and analysis
correspond to the grey, black, and dotted 
curves. A grey `$\diamond$' represent observations, 
and a `$+$'
corrections ($\vecX_a$) made by 3DVAR, 
both at assimilation time. 
}
\label{fig:illustrate} 
\end{figure}

Figure~\ref{fig:illustrate} clearly shows 
the result of this process
at times where observations 
(grey $\diamond$) are available,
3DVAR performs a correction given 
by equation (\ref{cost3d2}) to a value
referred to as analysis (black $+$), and
from which a segment of background states (dotted curve) is
initialized and run up to the next assimilation time. 
The analysis is the union of the 
background states for times
different from the assimilation times, 
and the corrected values obtained at
assimilation times. 

Given a model and a fixed set of observations, 
equation (\ref{eq:xa})
help us understand the effects of 
the weight factor in the resulting 
analysis $\vecX_a$, as it is detailed in the
following paragraphs. 

For $\wf<1$ 
or $\sigma_R < \sigma_B$, 
is interpreted as the case where we assumed
that the 
observational uncertainty is smaller 
than the model uncertainty, 
weight is given to the observations 
since small $\wf$ will allow the 
distance $(\vecX_a-\vecX_b)^2$ to
grow without making large contributions 
to the cost function. 
In contrast, having a $\wf>1$ will 
favor model states reflecting our
assumption that there is more uncertainty 
related to the observations
than that of the model, or $\sigma_R > \sigma_B$.
  
The next section presents and describes 
the results of our numerical experiments. 
We measure the ``quality'' of the obtained 
analyses using 3DVAR by varying 
values of the weight factor $\wf$ 
as well as a couple of sets of
observations with different noise levels. 

The correlation time was found to be approximately 2 
from simulation parameters (see Section \ref{sec:Strat}).
This determine the relevant time
scales of the convection features 
of our simulation and then we choose
the most appropriate assimilation time. 
Consequently, our choice of assimilation time is 
large enough to let the
oscillations propagate over the two-dimensional 
field but still small
enough to be able to capture the smaller scale dynamics.
In each 3DVAR experiment, data assimilation 
corrections are made each 10 time units 
(sound-travel-time units) and
found no fundamental difference when using
 assimilation times between
5 and 20 time units. 

\section{Results}
\label{sec:results}
We generate analyses using 3DVAR 
for each of the weight factors
$\wf=\{0,0.1,0.5,1,10\}$. 
For a fixed value of $\wf$ we generate 
an analysis which is the result
of assimilating one set of observations 
with either 1\% and 2\% noise
level for the same set of two initial field conditions.

\begin{figure}
\vbox{
\hskip-0.5cm
\includegraphics[width=9.5cm]{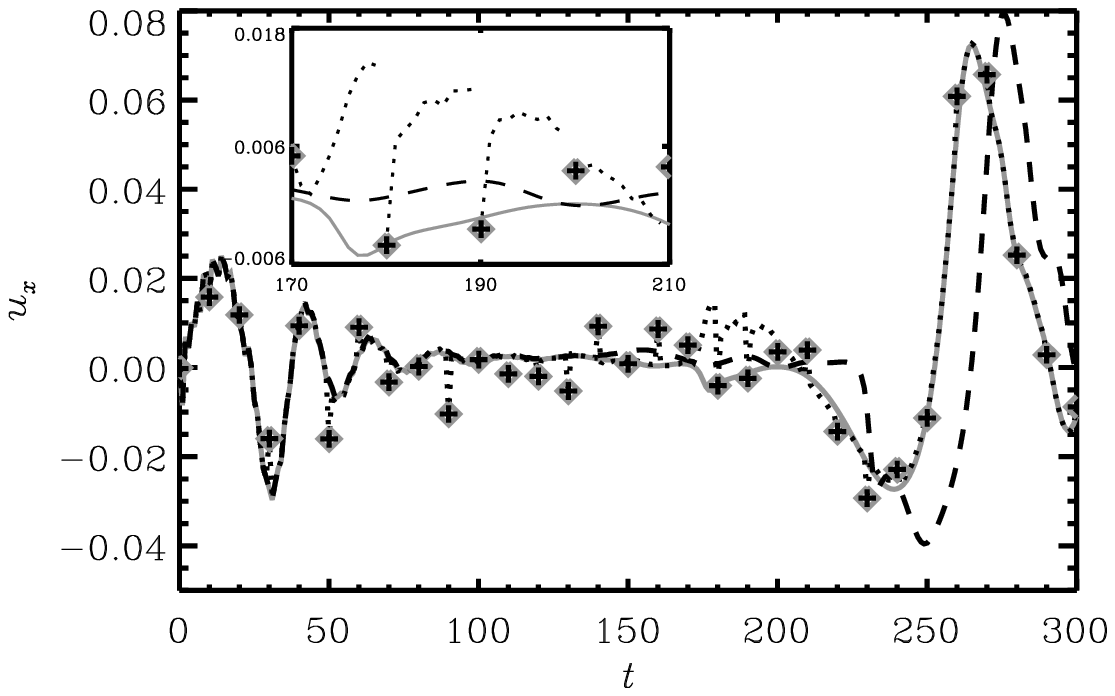}
\vskip-0.2cm
\hskip-0.5cm
\includegraphics[width=9.5cm]{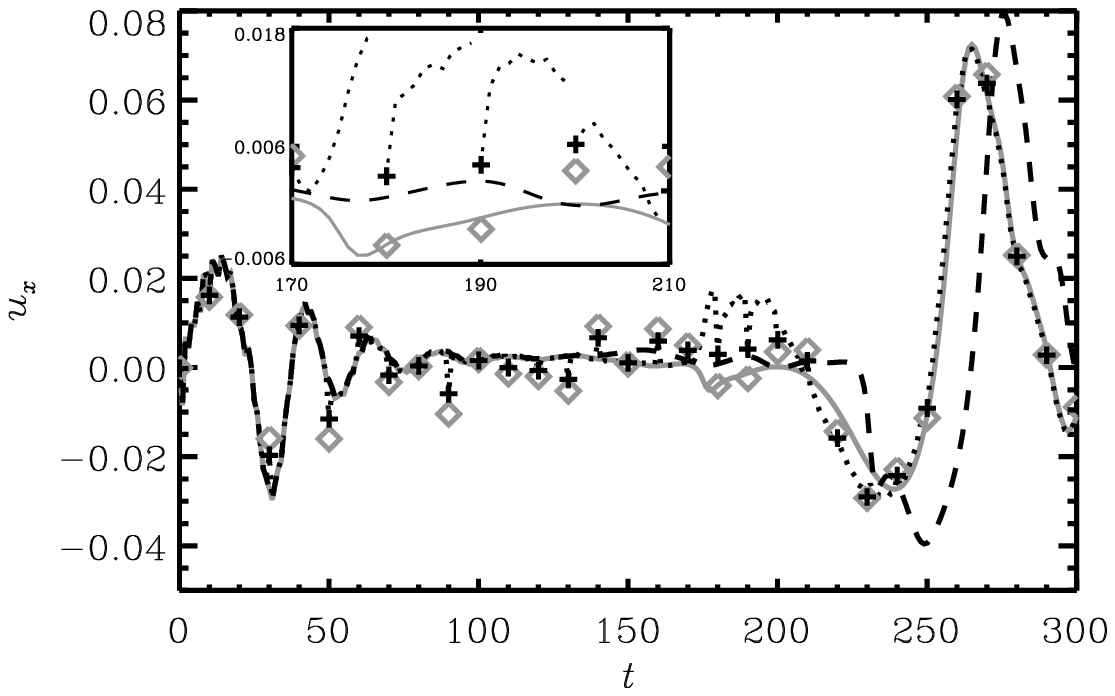}
}
\vskip-0.2cm
\caption{
  Data assimilation run over observations marked with grey `$\diamond$'
  with 1\% noise for $\wf=0$ (upper panel) and $\wf = 0.5$ (lower panel).
  Black `$+$' marked the 3DVAR correction at assimilation time.
  Grey, dotted, and
  dashed curves correspond to the control, analysis, and free trajectories,
  respectively. Both panels include zoom-ins for $t \in [170,210]$.} 
\label{fig:w0and05}
\end{figure}

\begin{figure}
\vbox{
\hskip-0.5cm
\includegraphics[width=9.5cm]{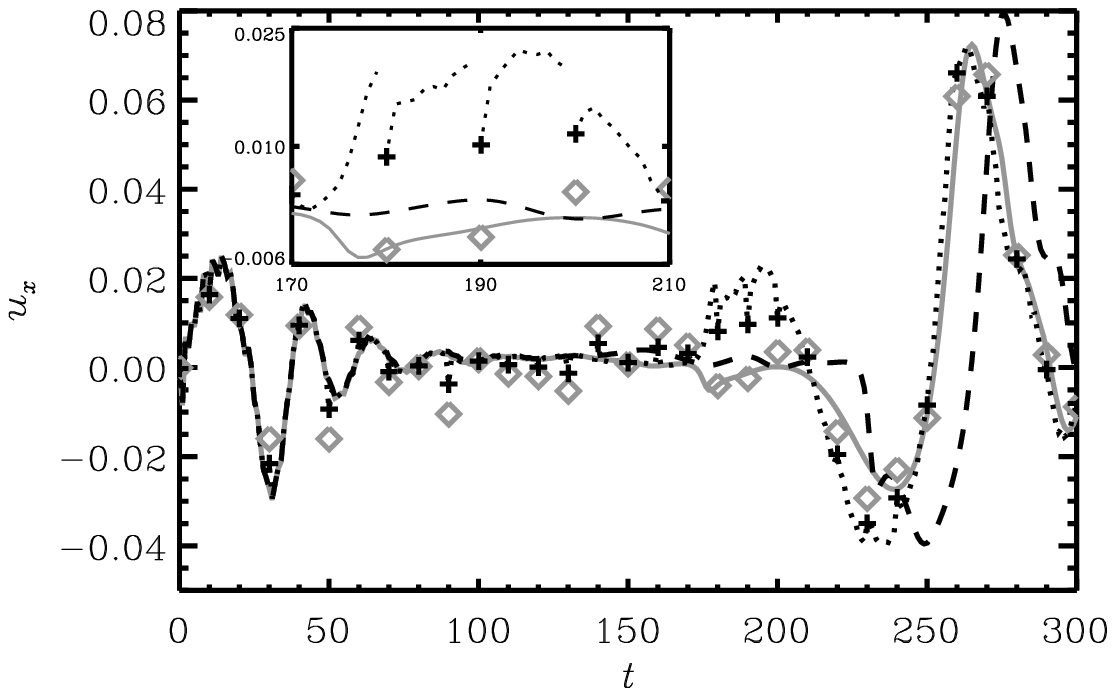}
\vskip-0.2cm
\hskip-0.5cm
\includegraphics[width=9.5cm]{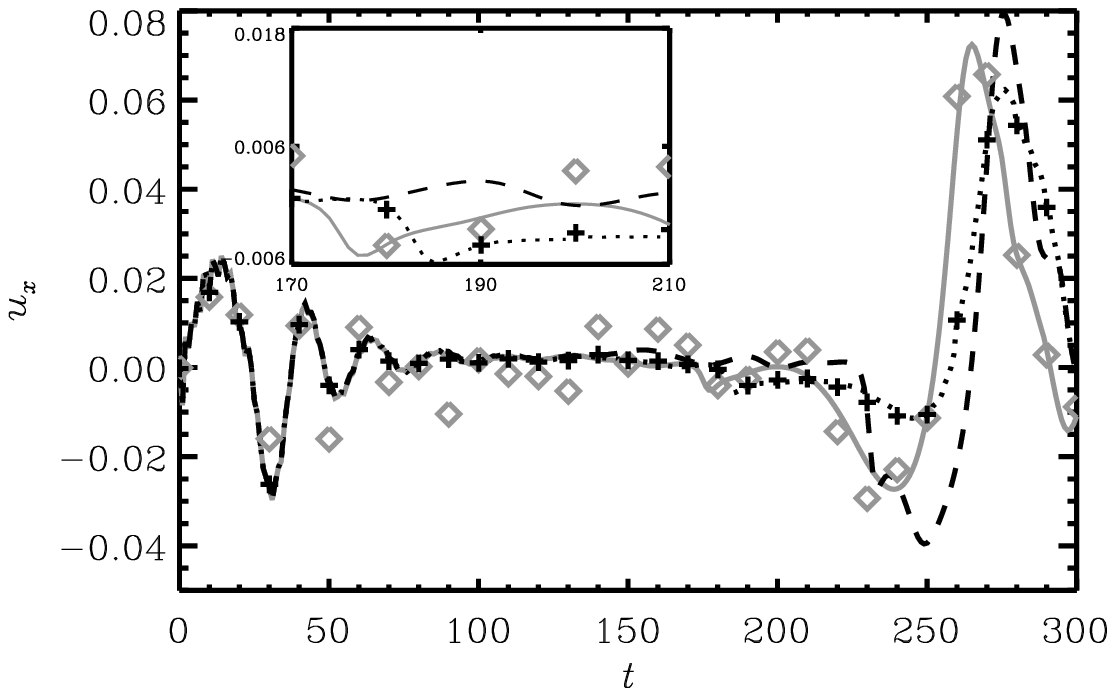}
}
\vskip-0.2cm
\caption{
  Data assimilation run over observations marked with grey `$\diamond$'
  with 1\% noise for $\wf=1$ (upper panel) and $\wf = 10$ (lower panel).
  Black `$+$' marked the 3DVAR correction at assimilation time.
  Grey, dotted, and
  dashed curves correspond to control, analysis, and free trajectories,
  respectively. Both panels include zoom-ins for $t \in [170,210]$.} 
\label{fig:w1and10}
\end{figure} 

The resulting horizontal velocity, $u_x$, 
at the midpoint of the upper
right quadrant of the two-dimensional domain, is plotted in
Figure~\ref{fig:w0and05} for
$\wf=0$ in the upper panel and $\wf=0.5$  in the lower panel, 
and in 
Figure~\ref{fig:w1and10} 
for $\wf=1$ in the upper panel and $\wf=10$ in the lower panel.

Note that grey and dashed lines are the same in all panels
since they represent the reference states of the system (control) and 
the corresponding free run of the model.

Observations are plotted with grey `$\diamond$' 
symbols. 
Black `$+$' marks are used for the corrections calculated at
assimilation time by minimizing expression (\ref{cost3d2}). 
Between assimilations, the analysis 
is composed by segments of model trajectories (dotted 
segments)
initialized at the corrected state `$+$' as seen in detail at 
all 
insets in 
Figs.~\ref{fig:w0and05} and \ref{fig:w1and10}.
From equation (\ref{eq:xa}) and these insets, we clearly illustrate
the amplitude of the correction made in each case for each
value of the weight factor $\wf$. 
This amplitude is measured by the gap between the `$+$' and the last
dot of the previous dotted segment (background states). 

For the trivial case of setting $\wf=0$, the 
first term in (\ref{cost3d2}) is neglected and the best estimate of
the current state of the system is given by $\vecX_a$= 
 $\vecY_0$ from in (\ref{eq:xa}), 
 and as seen in the upper panel of Figure~\ref{fig:w0and05}.
The correction (`$+$') 
is ``pulled'' from 
the background state
(dotted curve) to the
observation (`$\diamond$') at assimilation time.  
The background states are just segments of 
transient
trajectories initialized at the observations.  

For any other value of $\wf$ the starting 
point of the correction is
in between the observation (`$\diamond$') 
and the end of the previous background
states segment (dotted lines); see insets
in Figs.~\ref{fig:w0and05} and \ref{fig:w1and10}.

For $0 < \wf < 1$, the optimal value of the cost-function
(\ref{cost3d2}) is by a factor 
$1/\wf$ closer to the observations, $\vecY_o$,
than to the background state $\vecX_b$; see Eq.~(\ref{eq:xa}).
The lower panel of Fig.~\ref{fig:w0and05} 
shows results for $\wf=0.5$
where the corrections 
(`$+$') fall closer to the observations 
than to the end of the last segment of background states.
This figure shows that the analysis 
follows the control trajectory
closely (solid grey line). 
Note that even if the corrections are large, e.g.\
at $t=50$ or $t=90$, the analysis quickly 
relaxes to the control state.

In the case $\wf=1$, equal weights are given
to model states and observations.
The optimal value of $\vecX_a$ is the average of
$\vecY_0$ and $\vecX_b$, from (\ref{eq:xa}).
No preference is given
to any estimate and the midpoint is the 
optimal choice for $\vecX_a$
as seen in the upper panel of 
Figure~\ref{fig:w1and10}.
  
When $\wf>1$, the optimization of 
equation (\ref{cost3d2}) will favor
model states rather than observations 
as follows from equation
(\ref{eq:xa}).  
The analysis at assimilation times 
is by a factor $\wf$ closer
to the model states than to the observations.
In the lower panel of Fig.~\ref{fig:w1and10} 
we plot the resulting
trajectories for this case. 

For $\wf=10$, we observe from this 
plot and from results at other locations of
the two-dimensional domain that for 
$w>1$ the estimates of the original
state of the system are biased 
toward the background states.

When the model is linear and ideal, 
the analysis produced by 3DVAR
will be an unbiased estimate of 
the true state of the system. 
There, information about the 
system is only contained in the
experimental observations.
When the model is not ideal nor linear, 
information about the known unknown 
processes important to represent
the system of interest but not part of the model, 
could only be included via the $\wf$. 
In more general terms, through the 
covariance matrices $\matRR$ and
$\matBB$ which informed the 3DVAR 
procedure about model and
observational known uncertainties. 

In this simplified experiment, 
we can see the great importance of the
weight factor $\wf$ in 3DVAR. 
It points out how crucial the 
construction of the covariance
matrices $\matBB$ and $\matRR$ 
are for optimal results when using
variational approaches of data assimilation. 
One of the motivations for our choice 
of $\matRR$ and $\matBB$ to be
scalars is to set a baseline from 
which we can illustrate, in a simplified way,
the inner workings of 3DVAR.
It can be hard to see how the different components
interact to create a result when 
more sophisticated choices of $\matRR$ and
$\matBB$ are used. 

We would like to note that the zoom-in 
is chosen to be in the interval $t \in [170,190]$, 
in all panels of Figs. \ref{fig:w0and05} 
and \ref{fig:w1and10}, 
as an example of an interval where 3DVAR 
does not perform very well, and
systematically pulls the analysis 
away from the control
trajectory -- considered here as the original system trajectory. 
The reason for this behavior calls for further studies 
but it is worth noticing that around $t=200$ the performance
of the assimilation returns to its previous level. 
The case $\wf\ge 10$, presented in the lower panel of
figure 3, actually performs better during this 
interval. 

Consistently, we observe that the analysis
 is on average closer to the 
control trajectory than the observations
 for all values considered for  
$\wf\le 1$. 
As noted, exceptions are observed for larger values of
the weight factor and during the interval shown in the
insets of Figure 2 and 3.

Tables~\ref{tab:baselinevar} and~\ref{tab:variance} present 
several 
measures of variation of the 
output from the simulations of our twin
experiment.   
For each simulation, 
we calculate the variance of the distances
between the control and the 
observations (Table~\ref{tab:baselinevar}) 
or the control and the analysis
(Table~\ref{tab:variance}) over the data assimilation window.

Specifically, Table~\ref{tab:baselinevar} shows the variance 
of the distance between the control trajectory and the
free trajectory (first row), and the noisy 
observations for 1\% (second row) 
and 2\% (third row) noise levels at
the midpoint of the field (second column) denoted by
$\langle\langle(\delta{u^0_x})^2\rangle\rangle$, 
this column measures
the variability of the local behavior.
In addition, the averaged variance over the whole
vertical two-dimensional field for 
the first and the second half of
the assimilation window (third and forth column), denoted by
$\langle\langle(\delta^O {\mathbf u}_x)^2\rangle\rangle_T$ 
where the subscript $T=1,2$ refers to  
averages over the first or second half
of the assimilation window, $t\in[1,150]$ 
or $t\in[151,300]$ respectively.

Note that from the values in 
Table~\ref{tab:baselinevar}, the free run 
is one or two orders of 
magnitude further away from the original
state of the system (control) 
than the 1\% and 2\% noisy observations. 
Large difference 
between the second and third columns
reflect how free run is diverging
from the control run over two different time intervals.

\begin{table}
\centering
\begin{tabular}{crrr}
Noise Level & 
%\multicolumn{2}{|c|}{
$\langle(\delta u^O_x)^2\rangle$
%} 
& 
%\multicolumn{2}{|c|}{
$\displaystyle\langle\langle(\delta\mathbf{u}^O_x)^2\rangle\rangle_{1}$
%} 
&  
%\multicolumn{2}{|c}{
$\displaystyle\langle\langle(\delta\mathbf{u}^O_x)^2\rangle\rangle_{2}$
%}
\\
%(\%) & 
%\multicolumn{2}{c}{\tiny[$\times10^{-6}$]}&
%\multicolumn{2}{c}{\tiny[$\times10^{-6}$]}&
%\multicolumn{2}{c}{\tiny[$\times10^{-6}$]}\\ 
\hline
{free} & 1100 & 530 & 3000\\
1\% & 27 &  40  &  36\\
2\% & 110 & 140 & 140\\
\hline    
\end{tabular}
\caption{Measures of variability (in $10^{-6}$) for the distance
  between the control and
  observations with noise levels 1\%, 2\%, and the free trajectory.
  See text for notation description.} 
\label{tab:baselinevar}
\end{table}

Table~\ref{tab:baselinevar} is the baseline from which we
measure the performance of 
3DVAR when estimating the system state from
noisy observations or the free trajectory--or 0\% noise level--. 
Note that the variance of the noise is of the order of $10^{-6}$.
When assessing the performance of 3DVAR 
for different values of $\wf$
and noise levels, 
we look for variability measures between the control and the 
analysis lower than the levels 
set by the free run and the noisy
observations.  

All variance measures for the free trajectory are larger than the
corresponding variances values for all $\wf$ and noise levels. 
This means that performing 3DVAR data 
assimilation is more effective
at estimating the original state of the 
system than just using as
estimate the trajectory initialized 
with a very close initial field.
The variance of the distance between the two
initial velocity fields is $\approx2\times10^{-7}$.

Table \ref{tab:variance} shows the average 
variance of the distance between
control run and the analysis at 
the midpoint of the field (second
column) denoted by 
$\langle\langle(\delta{u^A_x})^2\rangle\rangle$, 
and the averaged variance over the whole
vertical two-dimensional field for 
the first and the second half of
the assimilation window (third and forth column), 
denoted by
$\langle\langle(\delta^A {\mathbf u}_x)^2\rangle\rangle_T$, 
for $T=1,2$, in analogy of the 
notation used in Table~\ref{tab:baselinevar}. 

From the values presented in both tables, 
we observe that for both noise levels 
studied and all cases with 
$\wf<10$, the variances of the distance 
between the control and the analysis are
smaller--on average--than the corresponding
observations with the same noise level.
This shows that on both local
and global scales, over the two-dimensional domain, 
3DVAR is effectively reducing 
noise and accounting for sensitivity to
initial conditions i.e.\ it is estimating a
value for the horizontal velocity closer 
to the original state of the
system (control) than to the trajectory 
generated using a guess of the
initial state of the system (free run).  

In contrast, for $\wf=10$ and both noise levels, 
the free and the analysis are--in
average--consistently farther away from the control 
than the original noisy observations
(corresponding values in Table~\ref{tab:baselinevar}) 
both locally and globally. 
This is also observed, for example 
at the specific location 
of the two-dimensional domain 
shown in figure \ref{fig:w0and05} and
\ref{fig:w1and10}. 

We note that during the time interval $t \in [170,210]$ 
(see insets of
Figs.~\ref{fig:w0and05} and~\ref{fig:w1and10})
this simplified version of 3DVAR under-performs
compared to other times. 
More sophisticated choices for $\matRR$ and $\matBB$, 
that inform more realistically 
the cost-function (\ref{cost3d2}) about spatial 
correlations over the field might
generate an improved and consistent performance.
When comparing in this way the 
variance of the distance between the control 
and the free run
 we write 
 $\langle\langle(\delta^F {\mathbf u}_x)^2\rangle\rangle_T$,
 for $T=1,2$.

\begin{table}
\centering
\begin{tabular}{cr@{.}lr@{.}lr@{.}l}
$\wf$ 
& \multicolumn{2}{|c|}{$\langle(\delta u^A_x)^2\rangle$} & 
\multicolumn{2}{|c|}{$\displaystyle\langle\langle(\delta\mathbf{u}^A_x)^2\rangle\rangle_{1}$} 
&  
\multicolumn{2}{|c}{$\displaystyle\langle\langle(\delta\mathbf{u}^A_x)^2\rangle\rangle_{2}$}\\ 
\hline
\multicolumn{7}{|c|}{Noise level 1\%} \\\hline
0 &  0&61 & 0&34  &  3&80 \\
0.1 &  0&67 &  0&35 &  4&40 \\
%1 & 3&30 & 4&10 & 130&00\\
%10 & 300&00 & 500&00 & 1600&00\\
1 & 3&30 & 4&10 & $\mathit{130}$&$\mathit{00}$\\
10 & $\mathit{300}$&$\mathit{00}$ & $\mathit{500}$&$\mathit{00}$ & $\mathit{1600}$&$\mathit{00}$\\
\hline
\multicolumn{7}{|c|}{Noise level 2\%} \\
\hline
0 & 2&40 & 1&30 & 19&00\\
0.1 & 2&70 & 1&40 & 22&00 \\
%1 & 33&00 & 9&80 & 840&00 \\
%10 & 950&00 & 430&00 & 2800&00 \\
1 & 33&00 & 9&80 & $\mathit{840}$&$\mathit{00}$ \\
10 & $\mathit{950}$&$\mathit{00}$ & $\mathit{430}$&$\mathit{00}$ & $\mathit{2800}$&$\mathit{00}$ \\

\hline
\end{tabular}
\caption{Measures of variability (in $10^{-6}$) for the distance between
  the control and the analysis for several values of the $\wf$ of the
  $x$--component of the velocity, $\mathbf u_x$. 
  The italic values are larger then the corresponding noise level.
  See text for notation descrition.}
\label{tab:variance}
\end{table}

\begin{figure}
\vbox{
\hskip-0.8cm
\includegraphics[width=9.5cm]{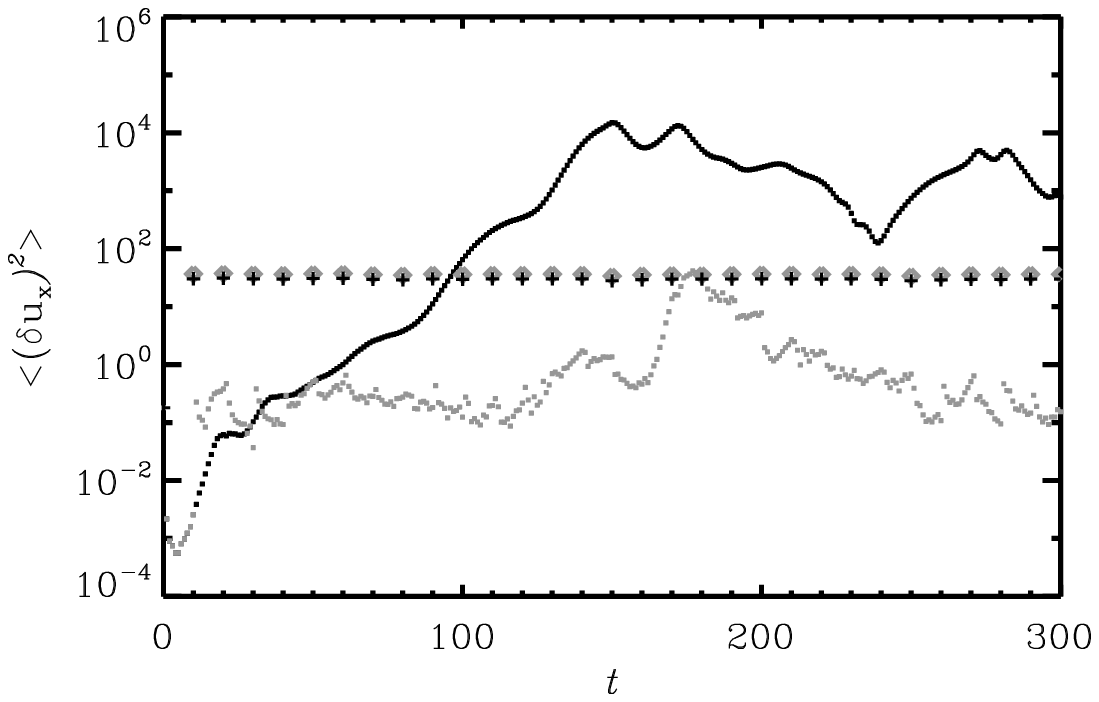}
\vskip-0.6cm
\hskip-0.8cm
\includegraphics[width=9.5cm]{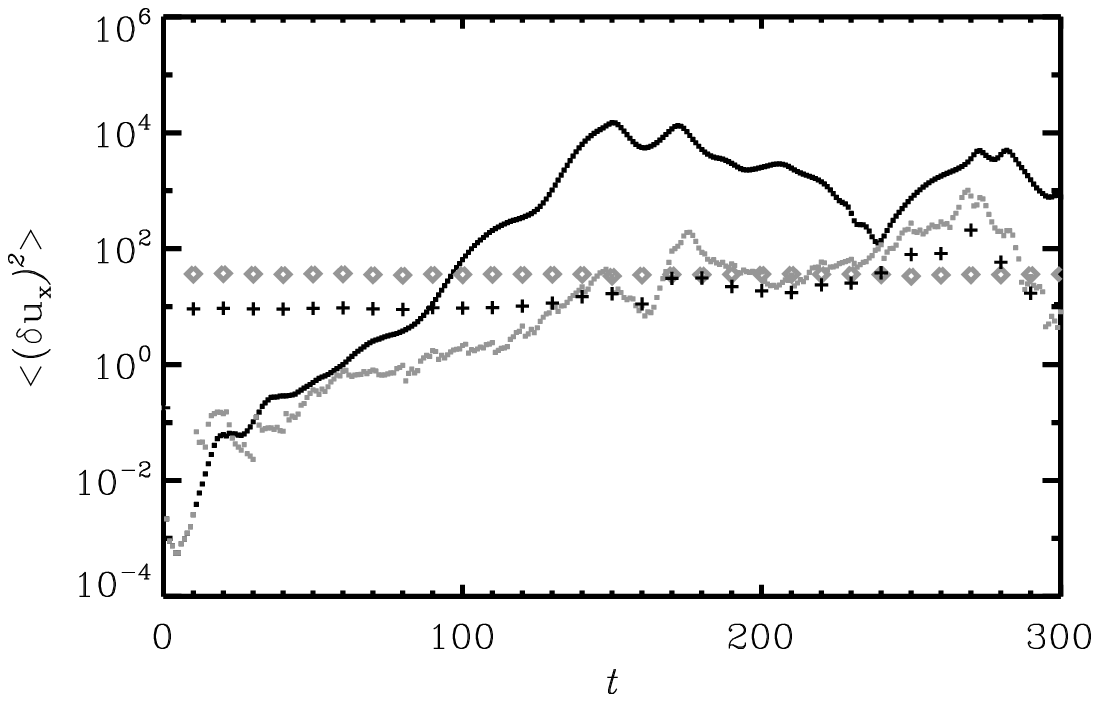}
%\vskip-0.5cm
\vskip-0.6cm
\hskip-0.8cm
\includegraphics[width=9.5cm]{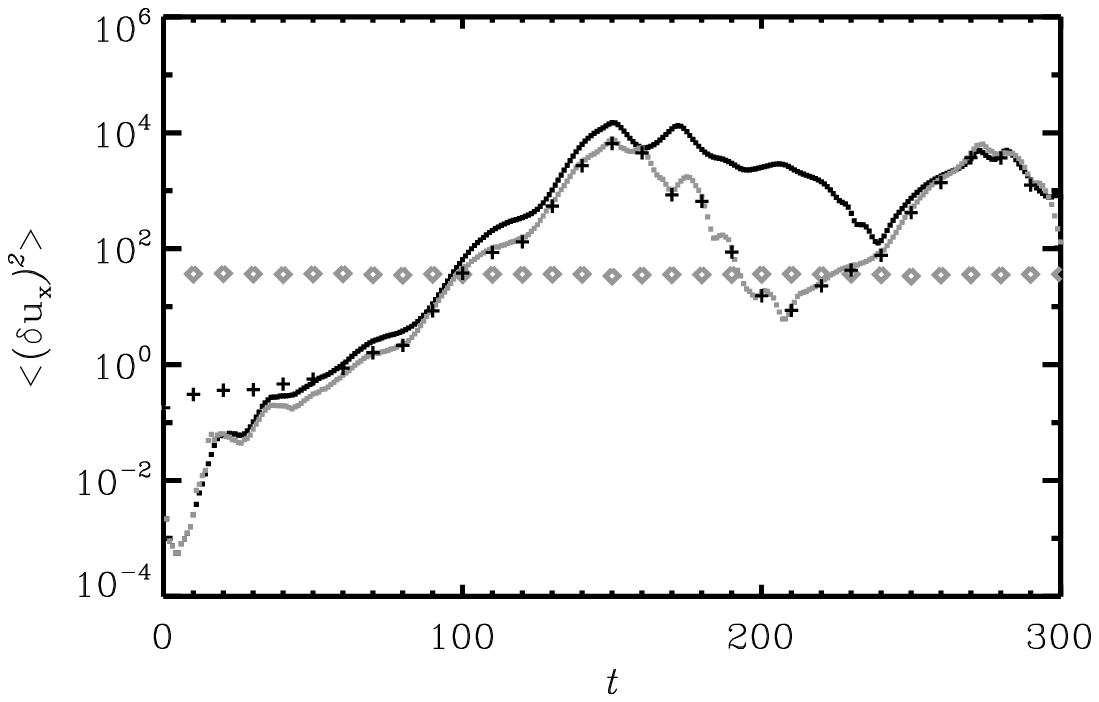}
}
\caption{Semi-logarithmic plot of the variance of the point-by-point
  distances of the two-dimensional vertical
slices between the control and the analysis in grey `$\bullet$' 
($\langle(\delta^A {\mathbf u}_x)^2\rangle$), the control and the
observations in grey `$\diamond$' 
($\langle(\delta^O {\mathbf u}_x)^2\rangle$), and
the control and the free run in black `$\bullet$' 
($\langle(\delta^F {\mathbf u}_x)^2\rangle$) for all $t\in[1,300]$.   
The black `$+$' corresponds to $\langle(\delta^A {\mathbf u}_x)^2\rangle$ 
at assimilation times, after the correction is made. 
All quantities are scaled by $10^6$.
}
\label{fig:tvar}
\end{figure}

Furthermore, in Figure~\ref{fig:tvar} we plot in a
semi-logarithmic scale the variance of the 
point-by-point distances of
the two-dimensional vertical
slices between the control and analysis (grey `$\bullet$'), 
the control and the
observations (grey `$\diamond$'), 
and the control and the free run
(black `$\bullet$') for all $t\in[1,300]$.   
The black `$+$' mark the variance of 
distances between the control
and analysis fields at assimilation times, 
i.e.\ when the corrections
from (\ref{eq:xa}) are made.

From top to bottom, plots correspond to $\wf=0.1,1,10$.  
Black dotted and grey $`\diamond'$ curves are 
the same for all panel in the
figure. 
The noise level is constant corresponding to 1\%.

It is important to note, that values plotted in 
Figure~\ref{fig:tvar}
are not the running variance of the distance 
between control trajectories
and the other relevant trajectories but the 
variance of the distances
between the 2D fields at each time step.

We observe in Fig.~\ref{fig:tvar} that for all values of
 $\wf$ and for $t \in [0,10]$ 
the analysis and free runs are good estimators 
of the control trajectory. 
In addition, for all $\wf$ and $t\in[20,100]$, 
the free run and the
analysis variances with respect to control, are
below the noise level, with the grey curve below the black curve,
showing that for 1\% noise level 3DVAR produces 
a better estimate than the free run.
Only for $w=0.1$ is this the case 
all values of $t\in[30,300]$. 
Note the increase in the error of the
estimate from $t\in[160,190]$. 
This is part of the interval 
shown in the insets for Figs.~\ref{fig:w0and05} 
and~\ref{fig:w1and10}).

Otherwise, the the error of the estimate
exceeds the noise-level for $t\in[170,300]$ 
for $\wf =1$, and  for
$t\in[190,230]$ for $\wf=10$.

Figure~\ref{fig:tvar} also shows how far apart--in 
average--the distance between the control and the free run grows as
time increases. 
The increase in the amplitude of the variance (the black curve) is up to
six or seven orders of magnitude
over the assimilation window with respect to the same distance at
small times.
As noted before, this exhibits 
the model' sensitivity to initial conditions and it is also a feature
observed for all values of the weight factor and noise levels.

In the top panel of the Figure~\ref{fig:tvar}, 
we again see evidence of how the 3DVAR
correction (black `$+$') 
lands closer to the observations and how the
background states start
to move towards to the control trajectory for lower values of the
weight factor, $\wf$. 
The
effect of $\wf<1$ in this case, 
is to ``pull'' the background states
back closer to the observations as 
the grey dotted curve is below the
corrections (`$+$'). 

We recall that the weight factor is $\wf=(\sigmaR/\sigmaB)^2$, 
and it is interpreted here as a measure of
the relative confidence given to either the observations and the model. 
High values of $\wf$ reflect higher trust in the observations
than in model representation of the system, and the opposite is
valid for low values of $\wf$.
Note that $\sigmaB$ relates to 
initial condition sensitivities than 
with model deficiencies in our simple setting. 
As commented earlier, the estimate is 
sensitive to the choices of
$\sigmaB$ and a more sophisticated choice
can include addition components
that might help alleviate some of the model
deficiencies.

In conclusion from the results of our experiment, 
in the particular
case where the covariance matrices are assumed 
to be scalar and for
small values of the weight factor $\wf$, 
the effect of trusting 
the observations more than 
the model states
($\sigma_R<\sigma_B$), provides a closer 
estimate of the original state of 
the system than just generating a
trajectory close enough in initial conditions. 
This means that 3DVAR is successful 
at finding an optimal state
estimator in the limit of small observational noise, 
model uncertainty
related only to sensitivity to initial conditions, 
and scalar
covariance matrices, $\matRR$ and $\matBB$. 
Our results strongly point out the importance of choosing more
sophisticated covariance matrices ($\matRR$ and $\matBB$) to better
inform the assimilation procedure about the known uncertainty sources
in a problem of interest. 

\section{Discussion}
\label{sec:discussion}

We have presented an idealized case where 
model and system are the same:
A computer simulation
is used both to generate synthetic
observations 
and as the model required for 
the data assimilation procedure.
In this way, we can assess how 
far/close the model state estimates are to
the true state of the system. 
The key is to have access to the true states 
that we can use to verify
and evaluate estimates
obtained using data assimilation. 

We used a simplified formulation of the 3DVAR 
data assimilation technique 
in terms of the weight factor: 
$\wf=(\sigma_R/\sigma_B)^2$, 
that defines the contribution 
of the model states -that contain 
propagated information from
previous observations- and the 
current observation to make
a state estimate. %

This formulation of 3DVAR used 
here is achieved by reducing the  
covariance matrices, $\matRR$ and $\matBB$, to scalars; 
and the observation operator, $\matHH$, to the identity. 
This selection corresponds to neglecting all spatial 
correlations between model states 
over the two-dimensional domain in addition to 
one-to-one correspondence
between system observables and model variables. 
In this way, we clearly separate the 
contribution of observations and model 
states to the estimated state, as seen 
in equation (\ref{cost3d2}) and (\ref{eq:xa}). 

It is less direct to see how the 
different components
interact to create an estimate of the 
original state of the system when more 
sophisticated choices of covariance matrices, 
 that represent uncertainties and spatial correlations. 
In that case we would have to
think about the optimal combination in analogy 
to equation (\ref{eq:xa}), in terms of a
generalization of the weight factor $\wf$ 
as a {\sl weight matrix}, 
${\matWW}=\matRR\matBB^{-1}$. 
In this analogy, model states and
observations will be projected by the matrices
 ${\matWW}[{\matWW + \mathbb I}]^{-1}$ 
and $[{\matWW + \mathbb I}]^{-1}$, respectively, 
to contribute to the analysis $\vecX_a$. 
Here, $\mathbb I$ is the identity matrix.

In general, we can say that to understand
the 3DVAR algorithm it is important to 
look at the weight factor, 
particularly in the limit where $\matWW$ 
is assumed to be the scalar $\wf$.

Consistently we observe that the error between the 
state estimate and the original state of the system 
is below the noise level when more weight is given
to the observations then to the model state. 
When the contribution from the model state is
larger than the contribution from the observations
we note that the error eventually becomes larger
than the noise level, see case $\wf=10$. 

We note in Figure \ref{fig:w0and05}, 
\ref{fig:w1and10} and \ref{fig:tvar}
that 3DVAR under-performs for $t\in[160,230]$
both locally and globally
for all values of $\wf$.
Further study of the simulation 
is needed to account for this
atypical behavior. 

Minute differences in initial conditions
generate different time evolution 
for the different runs,
as is expected for nonlinear systems.
This is illustrated by the black curves in Figure \ref{fig:tvar}
that present how the variance of the distance
between the two initial conditions grows over the 
time interval. 
It can also be seen in Figure \ref{fig:illustrate}
where the grey and black curves,
that started with close initial
condition, are very different 
at latter times.

On the other hand, large correction made by 3DVAR,
for example the black '+' at time 50 and 90
in Figure \ref{fig:w0and05},
does not appear to have a strong effect.
The model run that starts at these
far away estimates 
converges almost instantly back  
to the control at the same time.

These features, that is reminiscent  
of chaotic behavior,
can be understood in terms of
attracting sets where small changes 
in initial condition generate
a different time evolution on the attracting set. 
A large correction probably takes us 
outside the attracting set
and the solution rapidly falls 
back when the model is integrated forward.

Another interpretation is that the large 
correction takes us to states 
that are not consistent with 
conservation laws and other physical 
constraints.
The system would then 
rapidly be forced back 
onto a more physical state. 
This dual picture, using both physical and
mathematical intuition somewhat
clarifies this contradictory behavior. 

The 3DVAR methodology is optimal 
for linear models and
Gaussian distributed uncertainties 
\citep{lorenc1986}.
Very few models in astrophysics 
have these properties. 
The validity of variational 
methods outside the linear
or weakly nonlinear case is unclear
but if the assimilation is frequent enough
the behavior might be closer to linear.
Higher assimilation frequency during the
interval $t\in[170,230]$ might for example 
give a better result.

The indication of chaotic properties and 
attracting state space sets invites 
the use of other data assimilation methods that
explicitly take these properties into
account \citep{JS01,judd2008geometry}. 
They might be more applicable 
to non-linear astrophysical processes.
 
In this work we have used an ideal model, 
that is, the system and 
the model is one and the same. 
There will always be some limitation
to modeling of real system and this might
prove problematic. 
In general it is, in our mind, 
more important for models
used for prediction of real systems to be
reasonably realistic rather than just being close
to the observations.

For nonlinear astrophysical 
systems that operate on timescales 
from seconds to years, data assimilation will be 
of fundamental importance when quantitative agreement
between model and observations is to be assessed. 
The ultimate verification that a model is 
correct is its ability to make reliable prediction
and for this, data assimilation is necessary.

\section*{Acknowledgements}

We acknowledge the NORDITA data assimilation 
program of 2011 for
providing a stimulating scientific atmosphere.
This work was supported by the European 
Research Council under the
AstroDyn Research Project 227952.
We also like to thank Professor Spiegel and Dr Dobricic 
for their help in improving this manuscript.

\def\apj{ApJ}
\def\aap{A\&A}
\bibliography{ref}
\bibliographystyle{mn2e}

\end{document}